\documentclass[letterpaper]{article} 
\usepackage{aaai24}  
\usepackage{times}  
\usepackage{helvet}  
\usepackage{courier}  
\usepackage[hyphens]{url}  
\usepackage{graphicx} 
\usepackage{natbib}  
\usepackage{caption} 
\usepackage{algorithm}
\usepackage[draft]{fixme}
\usepackage[utf8]{inputenc}
\usepackage{dsfont}
\usepackage{subfiles} 
\usepackage{mathtools}
\usepackage[english]{babel}
\usepackage[utf8]{inputenc}
\usepackage{amsfonts, amsmath, amsthm, amssymb}
\usepackage{color}
\usepackage{natbib}
\usepackage{dsfont}
\usepackage{algpseudocode}
\usepackage{macros/macros}
\usepackage{macros/naveen_macros}
\usepackage{graphicx}
\usepackage{subcaption}
\usepackage{booktabs}
\usepackage{adjustbox}
\usepackage[shortlabels]{enumitem}

\theoremstyle{definition}
\newtheorem{definition}{Definition}
\newtheorem{theorem}{Theorem}

\newtheorem{assumption}{Assumption}

\usepackage{newfloat}
\usepackage{listings}
\DeclareCaptionStyle{ruled}{labelfont=normalfont,labelsep=colon,strut=off} 
\floatstyle{ruled}
\newfloat{listing}{tb}{lst}{}
\floatname{listing}{Listing}
\title{Observing Context Improves Disparity Estimation when Race is Unobserved}
\author{
    Kweku Kwegyir-Aggrey\textsuperscript{\rm 1}, 
    Naveen Durvasula\textsuperscript{\rm 2}, 
    Jennifer Wang\textsuperscript{\rm 1}, 
    Suresh Venkatasubramanian\textsuperscript{\rm 1}
}
\affiliations{
    \textsuperscript{\rm 1}Center for Tech Responsibility, Brown University, Providence, RI\\
    \textsuperscript{\rm 2}Columbia University, New York, NY\\
    kweku@brown.edu, nkd2126@columbia.edu, jennifer\_wang2@brown.edu, suresh\_venkatasubramanian@brown.edu
}

\usepackage{bibentry}

\FXRegisterAuthor{kk}{kkenv}{Kweku}
\FXRegisterAuthor{sv}{svenv}{Kweku}
\fxsetup{margin=True, inline=False}
\fxsetup{theme=color}
\definecolor{kkenv}{rgb}{0.8000,0.0000,0.0000}

\begin{document}

\maketitle

\begin{abstract}
In many domains, it is difficult to obtain the race data that is required to estimate racial disparity.  To address this problem, practitioners have adopted the use of proxy methods which predict race using non-protected covariates.  However, these proxies often yield biased estimates, especially for minority groups, limiting their real-world utility.  In this paper, we introduce two new contextual proxy models that advance existing methods by incorporating contextual features in order to improve race estimates. We show that these algorithms demonstrate significant performance improvements in estimating disparities on real-world home loan and voter data. We establish that achieving unbiased disparity estimates with contextual proxies relies on mean-consistency, a calibration-like condition. 
\end{abstract}

\section*{Introduction}
Algorithmic discrimination against a group is identified by determining if members of one group (usually the majority) group are more likely to receive some positive outcome than members of a minority or historically underrepresented group. These computations rely on knowledge of individual-level race data, which are then used to determine outcome disparities at the group level.  When these data are available, these computations are straightforward.  In many scenarios, however, individualized race data is difficult, and sometimes legally precarious,  to obtain \citep{andrus_what_2021}.

To address this challenge, many researchers have turned to proxy models, to help conduct disparity estimation \citep{elliott_new_2008}.  A proxy model offers predictions of an individual's race using non-protected features that are known to be highly correlated with race such as name, zip code, socioeconomic status, etc.  These predictions are then imputed onto an unattributed dataset, and then disparity is estimated with respect to the race predictions. 

The de-facto standard for proxy models is the Bayesian Improved Surname Geocoding (BISG) methodology \citep{imai_improving_2016}.  BISG  uses the United State's census tabulations of race by surname and geography, to predict the probability that an individual with a certain last name, living in a specific area, belongs to a racial group.

The use of BISG and other proxy models in disparity estimation has led to criticism from policymakers and fair practitioners who have found that proxy predicted race probabilities can lead to inaccurate disparity estimates \citep{adjaye-gbewonyo_using_2014, deluca_validating_2023}.  This estimation bias has been attributed to erroneous modeling assumptions, inaccuracies in the census data due to undercounting, and correlations between race and the outcome variable over which disparity is being estimated \citep{imai_addressing_2022}.  The exact nature of this estimation bias depends on the context where a proxy's predictions are operationalized.  For example, it has been reported that whether or not BISG overestimates or underestimates disparity, as well as the magnitude of estimation error, can depend on the racial composition of an area or on socioeconomic factors of the individuals whose race is being predicted \citep{argyle_misclassification_2024}.  

This observation highlights that disparity estimation is not one-size-fits-all and that context matters deeply when operationalizing a proxy method.  Unfortunately, many practitioners find that relying on a universal proxy method is their only solution, due to the many documented legal and procedural obstacles in collecting large amounts of attributed data within specific contexts \citep{kumar_equalizing_2022}. Context-specific data collection may also be redundant, as there are many instances in which the census's collected data is both accurate and well-calibrated enough to be useful in practice \citep{kenny_use_2021}. If we accept that this is true, it then begs the question: \emph{if a practitioner discovers that a proxy method does not perform well for their specific context, how should they improve the proxy's performance?}  

In this paper, we provide two new approaches to resolve that question.  We show that by using context-specific attributed data, in combination with existing census data, we can create what we call \emph{contextual proxy methods}, which lead to more accurate disparity estimation within their context.  We present two algorithms: \textbf{C}ontextual \textbf{B}ayesian \textbf{I}mproved \textbf{S}urname \textbf{G}eocoding (\textbf{cBISG}), and \textbf{M}achine Learning \textbf{I}mproved \textbf{C}ontextaul \textbf{S}urname \textbf{G}eocoding (\textbf{MICSG}).   Both approaches leverage BISG predictions and further modify them to produce improved race predictions.  We also present a new disparity estimator called the Bayes estimator and connect the bias of this estimator to a calibration-like property of the contextual proxy.\looseness=-1

\paragraph{Contributions. }  
\begin{enumerate}
    \item We provide two new contextual proxy algorithms, cBISG and MICSG which leverage context to provide improved race predictions.
    \item We present a new disparity estimator called the Bayes estimator that leverages contextual proxy models to estimate disparity.  We show that this estimator is unbiased if a proxy model satisfies an average conditional calibration notion called ``mean consistency'' \citep{jung_moment_2021}.
    \item We perform two large experiments using real-world data to demonstrate the success of our proposed approaches and our proposed estimator.  
\end{enumerate} 

\section*{Background}
In this section, we will introduce proxy models, describe how these models are used to identify disparities, and overview existing results that describe their known problems. The most widely used and studied proxy is BISG.  For this reason, BISG and its variants will be the main focus of this section.

\subsection*{What is a Proxy Model?}
A \textbf{proxy model} is a function that predicts an individual's race using some covariates that are highly correlated with race.  The covariates correlated with race are known as \textbf{proxy variables}.  Common proxy variables for race include name, income, location of residence, and political affiliation. 

Suppose that in our population there are $n$ individuals, each indexed with some $i \in [n]$. Every individual has some real valued, non-protected features $x_i \subseteq \mc X$ where $\mc X \subseteq \reals^{d}$, and a protected attribute denoted by the set $\mc R$.  Random variables over a set are notated with capital letters, e.g., the random variable associated with $\mc R$ is $R$.  For simplicity, we'll begin in the binary protected attribute setting by denoting race as a binary variable $\mc R = \{r_1,r_2\}$ where $r_1$ denotes the majority group, and $r_2$ denotes the minority group.   Our goal is to learn a randomized \textbf{proxy classifier} denoted $h:\mc X \rightarrow \mc R$ that predicts an individual's race.   
We model a proxy classifier as a randomized function because there is a nonzero probability that two individuals with the same feature vector belong to different races. For example, race is commonly predicted using last name and the area where one lives -- it is not uncommon for two people who live in the same area to share a last name but belong to different races.  Our main focus is the function that captures the randomness of $h$.  We call this function a \textbf{proxy model} and define it as $\rho_r(x) = \Pr[h(X) = r ~|~X = x]$ or as $\rho_r = \Pr[h(X) = r ~]$ when not conditioning on $x$. 
If a proxy model outputs the true race distribution for all $x$ such that the following holds
\begin{align}
\label{eq:proxy_cal}
    \rho_r(x) = \Pr[R = r~|~ X = x ], 
\end{align}
then we say that the proxy model is \textbf{calibrated}. 

We represent the \textbf{context} of individual $i$ as $y_i \in \mc Y$  where $\mc Y = \binaryset$ and $y_i=1$ denotes some outcome of interest e.g. receiving a loan.   An individual's context is determined by a \textbf{decision function} denoted $f:\mc X \rightarrow \mc Y$ where the decision function uses an individual's non-protected features to determine $y_i$.  The class of all possible decision functions (and their negation) is denoted by the set $f, \neg f \in \mc F$.  Note that the features used as input to a proxy may differ from the features used by a decision function; we decided not to make this distinction in our notation as we believe it to be a subtlety. 

\subsection*{Estimating Disparities via Proxies}
Before we provide a brief background on how to estimate disparity when protected attribute information is \textit{not} available, we will first describe the simpler setting of estimating disparity when protected attributes \textit{are} available. 

\subsubsection*{When Attributes Are Known. }
Informally, we say that a decision function is biased if it assigns $f(x)=1$ within one group more often than another. 
We formally represent the rate at which $f(x)=1$ is assigned within a group as the group conditional positive rate. 
\begin{definition}[Positive Rate]
\label{def: positive_rate}
The group conditional positive rate is computed  
\begin{align}
\label{eq: def_positive_rate}
    \mu_f(r) = \Pr[f(X) = 1 ~|~R = r]. 
\end{align}
\end{definition}

We measure \textbf{disparity} as the magnitude of the absolute difference between group conditional positive rates, i.e., as $\abs{\mu_f(r_1) - \mu_f(r_2)}$. When the group conditional positive rates are equal then $f$ is said to satisfy Demographic Parity, a common notion of algorithmic fairness.  
There are many ways to measure disparity.  Our results are sufficiently general such that they can easily be extended to many of these other notions by considering additional covariates when computing positive rates \citep{hardt_equality_2016}.  When protected attributes are known for each individual, i.e., samples from a joint distribution over $\mc X \times \mc R$ are available, estimating disparity is straightforward, since all quantities required to compute Eq (\ref{eq: def_positive_rate}) are known.

\subsubsection*{When Attributes Are Unknown. }
When protected attributes are unknown, the above quantities cannot be computed as we do not have individual level ground truth race data.  In these settings, a proxy model can be used to predict protected attributes, and then the bias is measured over the race predictions in hopes that the estimate is approximately equal to the actual underlying disparity \citep{imai_improving_2016}.  

There are many strategies for estimating disparity with a proxy. For example, a naive approach is to produce race classifications by thresholding a proxy model.  The thresholding approach deterministically assigns a race to an individual if $\rho_r(x) \geq \tau$ for some $\tau \in \binaryrange$. We do not study this approach in this work. \citet{dong_addressing_2024} show that this method can lead to systematic underreporting of disparity.

Another approach uses the outputs of a proxy model as weights in the disparity computation.  This estimator, known as the \textbf{weighted estimator} \citep{chen_fairness_2019}, is computed 
\begin{align}
\label{eq:weighted_estimator}
    \mu_f^W(r) = \frac{\sum_{i=1}^{n} \rho_r(x_i) * f(x_i)}{\sum_{i=1}^{n} \rho_r(x_i)}.
\end{align}

The main downside of this estimator is that it is proven to converge to a biased estimate for disparity, even with access to large amounts of data.  This bias is given by the following theorem.
\begin{theorem}[Theorem 3.1 of \citet{chen_fairness_2019}]
\label{thm:weighted_estimator_bias}
    \begin{align}
        \mu^{W}_f(r) - \mu_f(r) \overset{\text{a.s.}}{\rightarrow}  \frac{\expect[Cov(\indicator[ R= r], f(X) \mid  X )]}{\Pr[R = r]} 
    \end{align}
\end{theorem}

Thus, if there is a correlation between the output of the decision function and one's protected group, conditioned upon non-protected features, i.e., if race based discrimination exists in the output of the decision function, then the weighted estimator will necessarily produce biased disparity estimates.  For example, \citet{mccartan_estimating_2023} show that if black people are less likely than their white counterparts to receive a loan, then the weighted estimator will \textit{necessarily} under-report the black-white loan disparity.

\subsection*{Introducing Bayesian Improved Surname Geocoding}
\label{sec: introducing_bisg}
The Bayesian Improved Surname Geocoding (BISG) methodology is an application of Bayes Rule which can be used to predict an individual's race. In this section and what follows we will now assume race is a discrete set containing $\abs{\mc R} = K$ difference races. Using a surname $s \in \mc S$, and some geographic unit $g \in \mc G$, the BISG probability satisfies the following relation  
\begin{align}
    \Pr[R~|~S=s, G=g] \propto Pr[S = s ~|~ R] \Pr[R ~|~ G = g].
\end{align}

The two probabilities that are used to compute the BISG probability are $Pr[S~|~R]$, the probability that an individual carries a specific surname given their race, and $\Pr[R~|~G]$, the probability that an individual belongs to a certain race, given their geographic unit.   These probabilities are approximated from the United States Decennial Census, which tabulates a breakdown of both surname and geography by race.  This tabulation is made possible by representing race, surname, and geography, as finite sets.   Race is made finite by grouping individuals into one of the following categories: \{White, Black, Asian or Pacific Islander, Alaskan or Hawaiian Native, Mixed Race, or Hispanic\}.  Surnames are tallied over a finite list as census data only tracks names that appear at least 100 times.  The number of geographic units is also finite -- in our analysis, we work with census tracts (instead of other common choices like zip code, state, or county) of which there are 73,057 as per the 2010 census.  We chose census tracts because they are a relatively small geographic unit,  leading to more fine-grained estimates.

The computation of BISG probabilities relies on the following conditional independence assumption. 
\begin{assumption}[Conditional Independence of Surname and Geography given Race]
\label{assumption:bisg}
\begin{align}
    G \indep S ~|~ R
\end{align}
\end{assumption}

If this assumption is satisfied, we can derive the BISG probability by applying Bayes rule
\begin{align}
    \Pr[R~|~G,S] &\propto \Pr[S~|~R,G]\Pr[R~|~G] \\ 
    &= \Pr[S~|~R]\Pr[R~|~G]. 
\end{align}
This assumption, although widely accepted and mathematically convenient, is known to not always be suitable since it ignores the general wisdom that people tend to live closer with those to whom they are demographically similar.  For further discussion of the suitability of this assumption, we refer the reader to \citet{greengard_improved_2024, imai_addressing_2022}.  

This assumption aside, BISG is known to produce classifications that are well-calibrated and accurate at the population level \citep{deluca_validating_2023, kenny_use_2021}.  Unfortunately, many of these performance guarantees fall away when we condition on additional socioeconomic factors \citep{argyle_misclassification_2024}.  

\paragraph{Problems with Undercounting.} 
One known problem with BISG is that its data source, the United States Census, is not always accurate. 
The census often undercounts minority racial groups, and the census surname files omit many minority surnames because they appear at lower frequencies. 
These data issues contribute to inaccuracies within BISG, with a tendency to produce misclassification errors that are strongly correlated with race, and as we will see, context.

\section*{Our Method}
\label{sec: mlbisg}
To address some of the issues with BISG, and to provide a method for more accurate disparity estimation, we introduce our two algorithms, cBISG and MICSG. Our methods are based on the observation that the outcome variable over which disparity is being measured often provides meaningful information about an individual's race.  Thus, incorporating context into our predictions will lead to better race estimates, and subsequently, better disparity estimates.

Our first algorithm cBISG uses Bayesian inference to estimate race probabilities for every combination of context and geography.  We use these probabilities in combination with census name probabilities to obtain race estimates by surname, context, and geography, via a BISG-like formula.  Our method is novel, in that we use context in addition to surname and geography, to provide more accurate race probabilities. Our second algorithm MICSG is a supervised learning algorithm that uses the outcome variable as input to a supervised learning algorithm, in combination with the output of an existing proxy as a base model and other relevant covariates, to estimate race probabilities. Notably, MICSG only requires query access to a proxy model and can be used in combination with any proxy model (including BISG).

\subsection*{Incorporating Context into BISG}

\begin{figure*}[t!]
    \centering
    \begin{subfigure}[b]{0.3\textwidth}
        \centering
        \includegraphics[width=\textwidth]{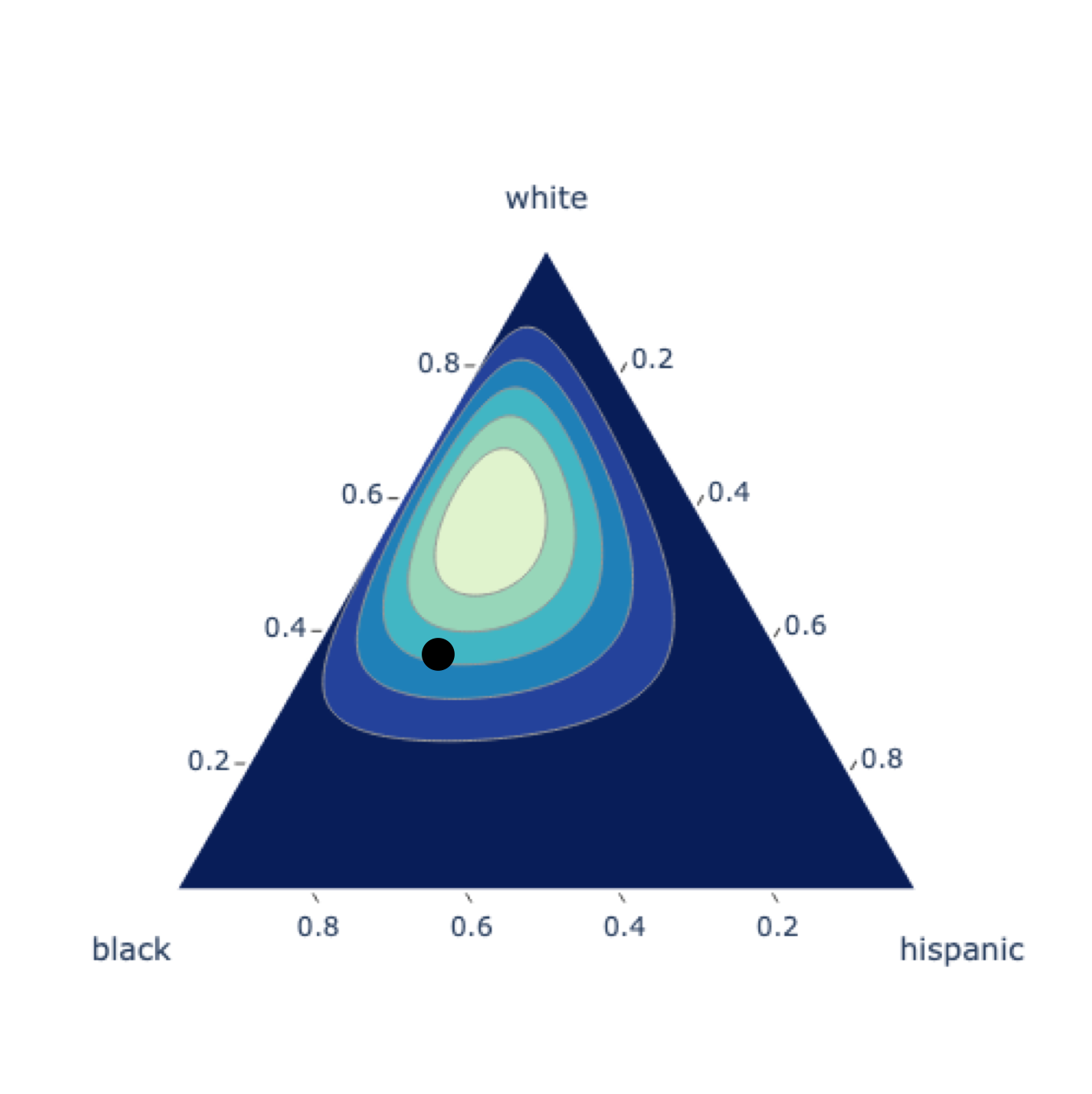}
        \caption{Census Prior}
        \label{fig:subfiga}
    \end{subfigure}
    \hfill
    \begin{subfigure}[b]{0.3\textwidth}
        \centering
        \includegraphics[width=\textwidth]{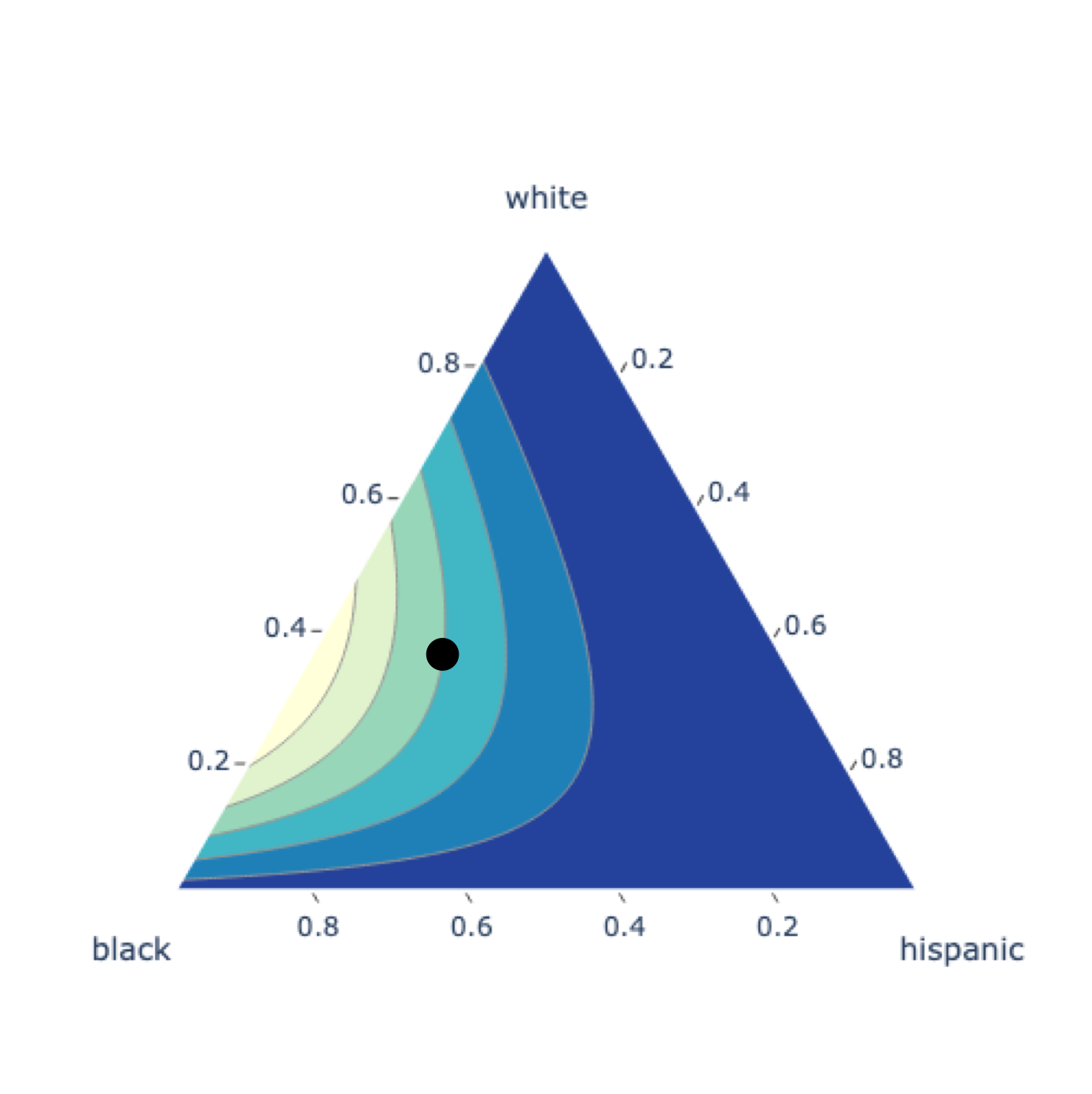}
        \caption{Posterior w/ Uniform Prior}
        \label{fig:subfigb}
    \end{subfigure}
    \hfill
    \begin{subfigure}[b]{0.3\textwidth}
        \centering
        \includegraphics[width=\textwidth]{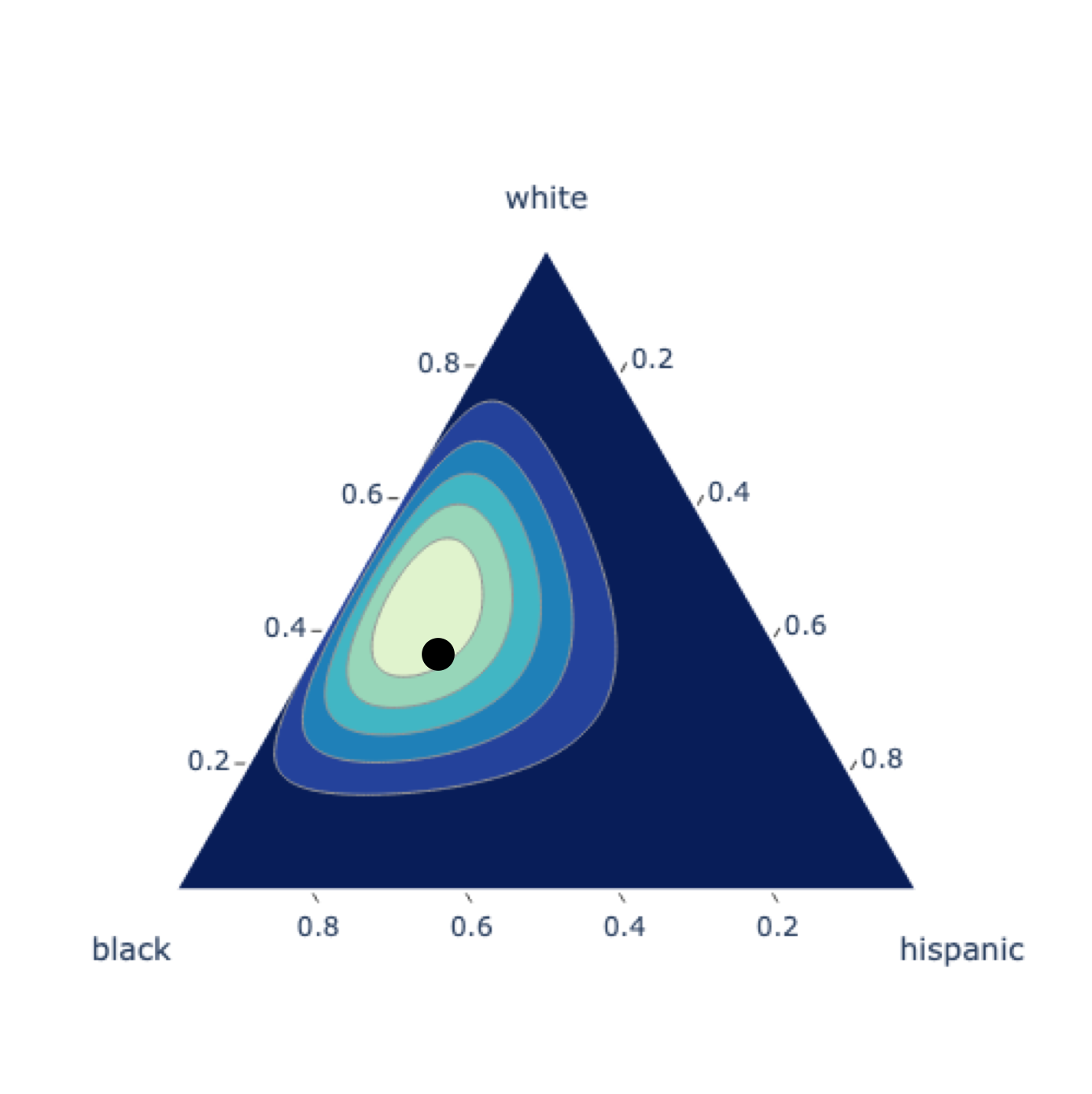}
        \caption{Posterior w/ $\eta$ adjusted Census Prior}
        \label{fig:subfigc}
    \end{subfigure}
    \caption{A toy example illustrating the effect of the $\eta$ hyperparameter on the posterior distribution (over distributions) for three races. Figure \ref{fig:subfiga} depicts the prior $\text{Dir}(5,3,2)$, which assigns higher mass to distributions that place a larger probability for observing white individuals, compared to other groups.  Now, suppose that in some supplemental data, we observe 2 White, 3 Black, and 1 Hispanic individual (the corresponding empirical distribution is indicated with the black dot).  Figure \ref{fig:subfigb} depicts the resulting posterior when $\eta = 0$, effectively assuming a uniform prior (rather than our census prior) thereby by assigning a higher likelihood to distributions that are majority black, as a consequence of the supplemental data.  Figure \ref{fig:subfigc} depicts the conjugate posterior for $\eta = 0.25$, where this posterior balances the census and supplemental race distributions.}
    \label{fig:subfigures}
\end{figure*}

The probability an individual belongs to a race $r$, given surname, context, and geography is 
\begin{align}
\label{eq: BISGY}
    \Pr[R= r ~|~G = g ,S = s,Y = y].
\end{align}

 To infer this probability from data using a BISG-like approach, we make the following assumption. 
\begin{assumption}[Conditional Independence of Surname, Context, and Geography given Race]
\label{assumption: CI_GS_RY}
\begin{align}
    S \indep \set{Y,G} | R
\end{align}
\end{assumption}

 This assumption is identical to Assumption \ref{assumption:bisg} in that we assume surname and geography are conditionally independent given race, however, we \textit{also} assume conditional independence of surname and race given the context variable. Suppose, for example, the context $Y$ denoted political party membership. This assumption asserts that an individual's political party does not provide any additional information about their last name if we know their race.

Under this assumption, we can derive an expression for Eq \ref{eq: BISGY}.  
By Bayes Rule,
\begin{align}
\label{eq:cbisg}
    \Pr[R~|~G,S,Y] &\propto \Pr[R~|~G,Y]\Pr[S~|~G, R, Y] \\ 
    \label{eq:cbisg2}
    &\propto \Pr[R~|~G,Y]\Pr[S~|~R]. 
\end{align}
where the second line follows from Assumption \ref{assumption: CI_GS_RY}. 
Hence, predicting contextual race probabilities reduces to estimating the above two quantities. 

To estimate $\Pr[S~|~R]$, like the standard BISG methodology, we can use census data.  Estimating $\Pr[R~|~G,Y]$ on the other hand cannot be done using census data since these data do not include knowledge of the context variable.    To infer this probability, we must use some attributed samples from a joint distribution over $\mc R \times \mc G \times \mc Y$. For example, a voter registration list could be a source of samples from this distribution. They include surnames, an individual's location of residence, and also their party affiliation.  In other settings, however, this data can be difficult to obtain.  In these cases, any proposed algorithm must be able to form good estimates with limited data. Our solution to this problem was to propose an algorithm that could leverage prior beliefs in the form of census data, and update these beliefs using some new observations that include context.  This is the basis for cBISG, which we present next. 

\subsection*{Contextual Bayesian Improved Surname Geocoding}
\begin{figure}
\centering
\noindent\fbox{\begin{minipage}{.95\columnwidth}
\paragraph{Algorithm: Contextual Bayesian Surname Geocoding.} 
\begin{enumerate}
    \item Infer $Pr(\boldsymbol{p^{(g)}} | \boldsymbol{G} = g, \boldsymbol{Y} = y)$ using supplemental data, census counts $\boldsymbol{C^{(g)}}$, and chosen $\eta^{(g)}$
    \begin{enumerate}
        \item Perform hyperparameter tuning on $\eta^{(g)}$ by selecting value which minimizes disparity on training data 
    \end{enumerate}
    \item Sample from this distribution to produce estimates $\Pr[R \mid G =g, Y=y]$
    \item Compute $\text{cBISG(r, g, s, y)}$ with Eq. \ref{eq:cbisg} using the probabilities computed in Step (2) 
\end{enumerate}
\end{minipage}}    
\caption{An overview of the cBISG algorithm for computing a contextual proxy.}
\vspace{-.4cm}
\end{figure}
Our algorithm \textbf{C}ontextual \textbf{B}ayesian \textbf{I}mproved \textbf{S}urname \textbf{G}eocoding (\textbf{cBISG}) can be summarized as the following: we use Bayesian inference to infer $\Pr[R~|~G = g,Y = y]$ for all pairs $\mc G \times \mc Y$.  Then, we estimate $\Pr[S \mid R]$ using census data, and multiply this probability by our estimation of $\Pr[R~|~G = g,Y = y]$.  The resulting product, as shown in Eq \ref{eq:cbisg2} will be the desired probability estimates $ \Pr[R= r ~|~G = g ,S = s,Y = y]$. 

To use cBISG, we assume access to a supplemental dataset consisting of $m$ tuples $(r_i, y_i, g_i, s_i)_{i=1}^{m}$. Because the algorithm is context-specific, we will produce a separate cBISG for each distinct value of $y$. We begin by modeling the number of individuals belonging to each race, that live in a geography $g$,  and have context $y$, as draws from a multinomial distribution parametrized by $\boldsymbol{p^{(g)}} = (p^{(g)}_{r_1}, p^{(g)}_{r_2},  \dots p^{(g)}_{r_m})$, where $\boldsymbol{p^{(g)}}$ are the true, but unknown proportions of each race living in $g$ for the chosen context, i.e, 
\begin{align}
\boldsymbol{n^{(g)}} \sim \text{Multinomial}(N^{(g)}, \boldsymbol{p^{(g)}})
\end{align}
and $N^{(g)} = \sum_{i=1}^{m} n_{r_i}^{(g)}$ is the number of people in the supplemental data living in $g$ with $y_i = y$.  We place a Dirichlet prior on the unknown race proportions in $g$,   
\begin{align}
\boldsymbol{p^{(g)}} \sim \text{Dir}(\eta^{(g)} * \boldsymbol{C^{(g)}})
\end{align}

where $\boldsymbol{C^{(g)}} = (C^{(g)}_{r_1}, C^{(g)}_{r_2},  \dots C^{(g)}_{r_m})$ are the census's counts of individuals of each race living in $g$ and $\eta^{(g)} \in \binaryrange$ is a concentration hyper-parameter that controls the influence of the census count on the prior distribution.  See Figure $\ref{fig:subfigures}$ for an illustration of the effect of hyperparameter. 
For any $g$, the full posterior distribution over the unknown race proportions is easily written in closed form since our Dirichlet prior and posterior are conjugate distributions given the multinomial likelihood
\begin{align}
    Pr(\boldsymbol{p^{(g)}} | \boldsymbol{G} = g, \boldsymbol{Y} = y) \propto \text{Dir}(\eta^{(g)}*\mathbf {C^{(g)}} + \mathbf{n^{(g)})}. 
\end{align}

To help interpret this posterior, we present the following observations:
\begin{enumerate}
    \item When $n^{(g)}$ = zero, i,e. no one in our supplemental data lives in $g$, then the posterior is governed exclusively by $\mathbf{C^{(g)}}$.  This means when supplemental data are few, our method will produce a probability estimate that is no worse than what BISG produces.
    \item The parameter $\eta^{(g)}$ can be set to balance the influence of the census data against that of the supplemental data when determining the posterior.  This is especially relevant in settings where $C_r^{(g)} >> n_r^{(g)}$ and the supplemental counts can be "drowned out" by the census counts.   
\end{enumerate}

\subsubsection{Is the Hyperparameter Necessary?}   The concentration parameter $\eta$ can have a profound effect on the success of our inference procedure.  While it can be set manually, we used a relatively lightweight hyperparameter optimization procedure.  Our approach was to select an $\eta$ for which the resulting posterior minimized disparity estimation error on training data. In general, we found that our method performed best with small $\eta$, indicating a relatively weak weight on our priors.  To elide this step, however, one can simply set $\eta^{({g})} = 0$ for all $g$.

\subsection*{Predicting Race under Query Access to a Proxy Model with MICSG}
\label{sec:MICSG}

Suppose that in addition to conditioning on some outcome variable, we'd also like to include some relevant covariates in our race predictions.  Doing so in the cBISG scheme is difficult since we would need to learn a density over a joint distribution involving these covariates. If these covariates are high dimensional or continuous, learning such a density will require large amounts of data which may not be available in this setting.

A known solution to this problem is to feed the predictions of a proxy method, in addition to some other covariates,  as input into a supervised learning algorithm, and to predict race by learning \citep{ decter-frain_how_2022}.  This has the benefit of only requiring query access to the proxy model and allows for the inclusion of additional features to help improve predictions.  

Our MICSG algorithm makes one simple, but powerful, modification to this approach.  We treat the outcome variable as a feature when learning to predict race.  The intuition behind this modification stems from the fact that an individual's outcome is a meaningful proxy for race. For example, for mortgage data, we found that loan denial was highly correlated with being black.  Similarly, in voter registration datasets, being white is highly correlated with being Republican. Hence, the outcome variable for which disparity is being estimated can be treated as a proxy variable due to its high correlation with race.  Including outcome as a feature increases the predictive accuracy of our proxy. \footnote{We also remark that including the outcome variable as an input to our model is safe because we are not predicting outcomes -- we are predicting race distributions.}

\begin{figure}
    \centering
\noindent\fbox{\begin{minipage}{.95\columnwidth}
\paragraph{Algorithm: Machine Learning Improved Contextual Surname Geocoding.} 
\begin{enumerate}
    \item Query a base proxy model to obtain race predictions $\boldsymbol{\rho(x_i)} = (\rho_{r_1}(x_i) \dots  \rho_{r_K}(x_i))$ for each $x_i$ in the supplemental data. 
    \item Concatenate base proxy predictions with context variable $y$ and additional covariates $\boldsymbol{z_i} \in \mc X$ to form new feature vector $\boldsymbol{x_i^*} = (\boldsymbol{\rho(x_i)}, \boldsymbol{z_i}, y_i )$
    \item Apply a supervised learning algorithmic on the pairs $(\{\boldsymbol{x_i^*} , r_i)\}_{i=1}^{m}$ to obtain MICSG predictions 
\end{enumerate}
\end{minipage}}    \caption{An overview of the MICSG algorithm for computing a contextual proxy.}
    \label{fig:enter-label}
\end{figure}

Because we are learning a distribution over classes, we recommend learning algorithms that are well-suited for multiclass prediction, namely multinomial logistic regression and gradient boosting. 
\section*{Mean Consistent Proxies and Unbiased Disparity Estimates}
\label{sec: mean_consistency_proxies}
In this section, we prove one of the key advantages of contextual proxy models -- they can be used to attain unbiased estimates of an underlying disparity. We contrast this with other estimators, such as the weighted estimator in Eq. (\ref{eq:weighted_estimator}) which is known to be biased (see Theorem $\ref{thm:weighted_estimator_bias}$).  

To complete this analysis, we must define some new notation.  Let $\omega_r^f: \mc X \times \mc Y \rightarrow \binaryrange$ be a \textbf{contextual proxy model} where
\begin{align*}
    \omega^f_r(x, y) = \Pr[h(X) = r~| ~f(X) = y, X=x].
\end{align*}  This will be the output of cBISG or MICSG. If our proxy model outputs the true randomness of $h$ conditioned on context, then we will call the proxy model mean consistent.  
\begin{definition}[$\epsilon$-Mean Consistency]
Fix a context $y \in \mc Y$. A contextual proxy model is $\epsilon$-mean consistent for a context $y$ and with respect to some $f \in \mc F$ if 
the following holds 
\begin{align}
\abs{\expect_{X}\bra{\omega^f_r(x,y)} - \Pr[R = r ~|~f(X) = y]} \leq \epsilon.
\end{align}  
When $\epsilon=0$, we will say that the proxy model is mean consistent.
\end{definition}

The estimator that we use to compute disparities using a context-aware proxy model is called the \textbf{Bayes estimator} and is defined
\begin{align}
\label{eq: bayes_estimator}
{\mu}_f^B(r) = \frac{\sum_{i=1}^n \omega^f_r(x_i, 1) \cdot \sum_{i=1}^n \indicator[f(x_i) = 1]}{\sum_{y \in \mc Y}\parens{\sum_{i=1}^n \omega^f_r(x_i, y) \cdot \sum_{i=1}^n \indicator[f(x_i) = y]}}.
\end{align}

Mean consistency is a powerful tool for analysis in our disparity estimation setting as we prove that mean consistent proxies yield unbiased disparity estimates using the Bayes estimator.  We prove this via the following theorem.
\begin{theorem}
\label{thm: unbiased}
If $\omega^f_r(x,y)$ is mean consistent for all $y \in \mc Y$, then for all $r \in \mc R$ the Bayes estimator is an unbiased estimator for $\mu_f({r})$, i.e., ${\mu}_f^B(r) \rightarrow \mu_f({r})$ as $n\rightarrow \infty$.
\end{theorem}

The key idea that supports this result, is the fact that given enough data and a mean consistent contextual proxy, the Bayes estimator converges to
\begin{align*}
\frac{\Pr[R =r ~|~ f(X)=1]\Pr[f(X)=1]}{\sum_{y \in \mc Y}\Pr[R =r ~|~ f(X)=y]\Pr[f(X)=y]} 
\end{align*}
which is exactly $\mu_f(r)$ by Bayes rule. All proofs can be found in the appendix. 

\begin{figure*}[t!] 
  \includegraphics[width=\textwidth]{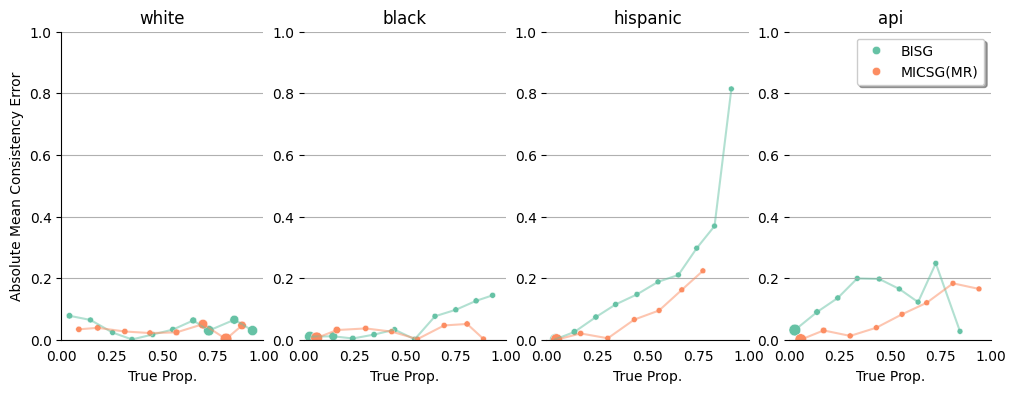} %
  \caption{We show the mean consistency violations for BISG and MICSG.  The x-axis denotes some true proportion of individuals per racial group, in some geography, who received loans. The size of each dot denotes the size of the bins indicated on the x-axis. The y-axis denotes the mean consistency violation of the corresponding proxy model. Being close to the horizontal $x=0$ line indicates good performance. MICSG leads to smaller violations compared to BISG across all groups. }
  \label{fig:hmda_consistency_vio}
\end{figure*}

In summary, Theorem \ref{thm: unbiased} provides a pathway towards unbiased disparity estimation using contextual proxies.  A contextual proxy that satisfies mean consistency paired with the Bayes estimator will result in unbiased estimation.  Additionally, we highlight to the reader that the Bayes estimator is not only usable with contextual proxies. Ostensibly, a practitioner could take a non-contextual proxy, acquire some context through a decision function, and then evaluate the Bayes estimator that way.  The disadvantage of that approach is that non-contextual proxies often satisfy marginal calibration guarantees but are not mean consistent, i.e., they are less calibrated when conditioned on context (see Figure \ref{fig:hmda_consistency_vio}).  This means, that a Bayes estimator paired with a non-contextual proxy will not lead to unbiased estimation.

\subsection*{The Effect of Mean (In)-Consistency}
\label{sec: mean_inconsistency}
While we have just established that mean consistent proxies are necessary for unbiased disparity estimation via the Bayes estimator, it perhaps even more important we characterize the effect of proxies that are \emph{not} mean consistent on disparity estimation.  In this section, we develop an analysis of proxies that do not satisfy $\epsilon=0$ mean consistency.  The goal of this analysis will be to examine the difference between the fraction of individuals belonging to race $r$ conditional on the $y=1$ context denoted $\phi_r \coloneqq \Pr[R = r ~|~ f(X) = 1]$
and the average value of a contextual proxy for the $y=1$ context denoted $\Bar{\omega}^f_r \coloneqq \expect_{X}\bra{\omega^f_r(x,1)}$. 
We will show that the bias of our disparity estimation depends on the absolute difference between these terms $\abs{\bar{\omega}_r^f - \phi_r}$, which we call the \textbf{mean consistency violation}.

The main result of this section is a pair of theorems that shows the relationship between mean consistency violations and $\epsilon$-biased disparity estimates using the Bayes estimator.  Before presenting these results, we define the following two quantities: the marginal race proportion $\theta_r \coloneqq \Pr[R = r]$, and $\nu_f \coloneqq \Pr[f(X) = 1]$, the rate that $f(X)=1$ averaged over the entire data distribution.  Additionally, we will say that an estimate is $\epsilon$-biased if it is $\epsilon$ far from some intended quantity (in absolute value).

Now, we will present our first theorem which states that for a fixed $f$, you can obtain $\epsilon$ biased estimates for $\mu_f(r)$ if our proxy model is mean consistent up to a factor proportional to $\epsilon$. 
\begin{theorem} \label{thm:mean consistent_to_bias}
Fix $\epsilon \in \binaryrange$ and let $f \in \mc F$ be any decision function. If $\rho_r$ is a $\gamma$-biased estimate for $\theta_r$ and $\omega^f_r$ is $\paa{\frac{\epsilon\theta_r}{\nu_f} - \frac{\bar{\omega}^f_r\gamma}{\theta_r - \gamma}}$-mean consistent
then, $\mu^{B}_f(r)$ is an $\epsilon$-biased estimate for $\mu_f(r)$. 
\end{theorem}

This theorem helps elucidate the connection between mean consistency violations and the bias of the Bayes estimator.  The most important thing to note is that the strength of this connection depends on the data distribution, $f$, and $h$.   In other words, the bias that we observe from the Bayes estimator depends on quantities that are not fixed between disparity estimation instances. For example, suppose a practitioner's goal is to compute estimates using the Bayes estimator that are biased up to some small $\epsilon$.  If $\nu_f$ is large (close to 1) and $\theta_r$ (close to zero) is small, then the proxy model must admit a relatively small mean consistency violation for the bias of $\mu^B_f(r)$ to remain small, as larger violations will increase the magnitude of estimator's bias quickly. On the other hand, if $\nu_f$ was small (close to 0) and $\theta_r$ was large (close to 1), then we would still be able to attain $\epsilon$ biased disparity estimates, but with a proxy model that produces a much larger consistency violation compared to the first scenario.

We can go a step further, and prove that attaining mean consistency is equivalent to attaining unbiased disparity estimates via the Bayes estimator.  With Theorem \ref{thm:mean consistent_to_bias}, we have shown that mean consistency implies unbiased disparity estimation.  The following theorem shows this implication in the other direction; that unbiased disparity estimation implies mean consistency.  
\begin{theorem}\label{thm:bias_to_consistent}
If for any $f \in \mc F$ it holds that $\mu_f^B(r)$ is an $\epsilon$-biased estimate for $\mu_f(r)$ then $\omega^f_r$ is $\paa{ \frac{\epsilon \theta_r}{\eta_f} + \gamma \frac{ \mu^B_f(r)}{\eta_f}}$ mean consistent.
\end{theorem}

These two theorems together demonstrate that mean consistency is both necessary and sufficient to attain disparity unbiased disparity estimates.  This result is of practical use for practitioners, as it locates the source of poor disparity estimates (in a contextual setting), in the proxy itself.  We contrast this to the weighted estimator whose bias term is data-dependent.  By locating the source of estimation error as a property of the proxy model, this result enables practitioners to better diagnose the source of bias in their specific context. 

\section*{Related Work}
What many consider as the original proxy model, BISG was first presented by \citet{elliott_new_2008} to help improve race data in healthcare records.  Since then, BISG has been applied in a variety of domains such as \citep{imai_improving_2016} including fair lending \citep{kumar_equalizing_2022}, political redistricting \citep{deluca_validating_2023}, and transportation \citep{sartin_facilitating_2021}.  Many variants of BISG predate ours. One of the most prominent is Bayesian Improved First Name Surname Geocoding (BIFSG) which uses first name supplements to improve the name coverage of BISG \citep{voicu_using_2018}.  \citet{imai_addressing_2022} present another version of BISG entitled fully Bayesian Improved Surname Geocoding (fBISG) relying on Bayesian inference to address issues in census undercounting. 

Some non-Bayesian approaches to race prediction by proxy include \citep{decter-frain_how_2022} who show that supervised machine learning based methods sometimes outperform BISG at individual prediction. \citet{greengard_improved_2024} also improve on BISG using supplemental voter records, and a rankings-based approach. Our MICSG approach is most similar to concurrent work from \citet{argyle_misclassification_2024} who show that a random forest trained on BISG probabilities and some socioeconomic covariates can help reduce misclassification rates for individual race prediction on voter files.  The key difference between their approach and our work is that our approach is agnostic to both the choice in the learning algorithm and base proxy.  This means that we can use learning algorithms like gradient boosting or multinomial regression to produce race predictions (not only random forests), in tandem with proxy predictors other than BISG, like BIFSG, fBISG, or even non-bayesian approaches.

A separate line of work studies how BISG and related techniques can inject bias into disparity estimation.  One of the main results in this aspect comes from \citet{chen_fairness_2019} who provide data-dependent bounds on the bias of the weighted estimator.  \citet{kallus_assessing_2022} also provide bounds on disparity estimates by proxy, using a data combination approach, while \citet{prost_measuring_2021} analyze disparity estimation under various data models and independence assumptions.   The weak relationship between disparity estimation and proxy model accuracy is established in \citep{awasthi_evaluating_2021} and \citep{zhu_weak_2023} who both contend that accurate proxies do not necessarily lead to accurate disparity estimates. \citet{mccartan_estimating_2023} contribute a Bayesian approach to improving disparity estimates by leveraging BISG probabilities and a user-specified model to estimate outcome probabilities conditioned given race.  While their approach is similar to ours, a key difference is that we focus on improving race predictions before estimating some disparity, whereas \citet{mccartan_estimating_2023} estimate disparities directly.  In \citet{fabris_measuring_2023}, disparity is estimated by using quantification techniques to estimate the prevalence of an outcome within each racial group.

Our work connects with recent topics uncertainty quantification, in that we use a calibration-like-notion called mean-consistency \citep{jung_moment_2021} to attain our theoretical results.  These contributions build on results from \citet{diana_multiaccurate_2022} who are the first to connect disparity estimation via proxy with conditional model performance measures, namely multiaccuracy \citep{kim_multiaccuracy_2019}.  In their work, however, they focus on learning multiaccurate proxy models rather than on disparity estimation. 

\begin{figure*}[t!] 
  \includegraphics[width=\textwidth]{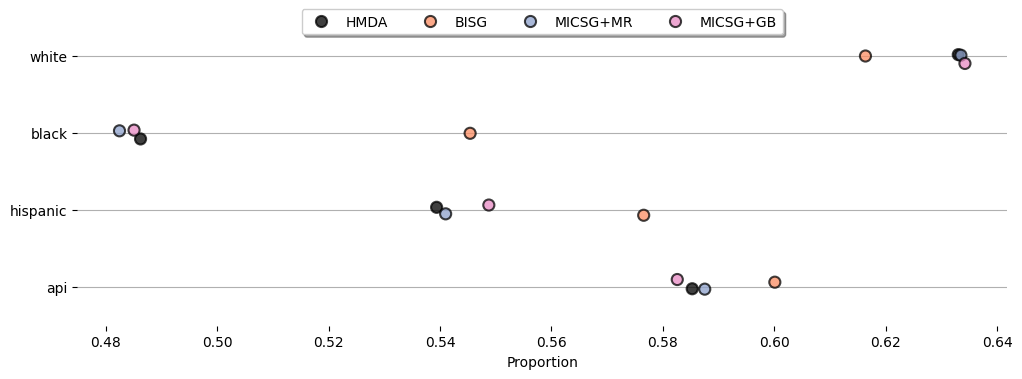} %
  \caption{The loan approval rate per racial group is given by the x-axis.  A dot closer to the HMDA dot implies more accurate disparity estimation. MICSG variants outperform BISG across all groups.} 
  \label{fig:hmda_MICSG_vs_bisg}
\end{figure*}

\section*{Experiments}
\label{sec:experiments}
In this section, we provide two case studies to demonstrate the success of our proposed approaches. In the first case study, we examine home mortgage approval rates by race and show that using MICSG, we can produce \emph{highly} accurate estimates of the disparities in mortgage approval between groups.  In the second case study, we look at the racial composition of political parties in North Carolina. We use records from the North Carolina voter registration database to train cBISG, using political party affiliation as the context variable.  Our experiments show that cBISG leads to significantly better estimates of the racial composition of each party compared to the BISG methodology and related approaches.

\begin{figure*}[t!] 
  \includegraphics[width=\textwidth]{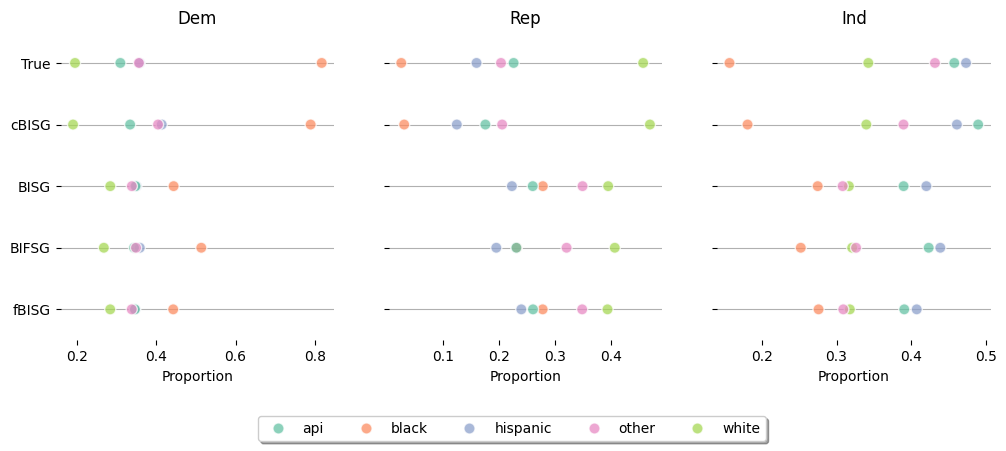} %
  \caption{The proportion of individuals across three political parties who belong to one of five racial groups in North Carolina. Each plot is replicated identically across the three political parties listed at the top of the figure. The ``True'' legend indicates the actual reported percentages of party membership, conditioned on racial group. A line of dots which appears similar to the ``True'' line, indicates good performance. cBISG significantly outperforms the BISG methodology and its variants.}
  \label{fig:twocolumn}
\end{figure*}

\subsection*{Case Study: Racial Disparities in Mortgage Financing}
In 1975, congress passed the Home Mortgage Disclosure Act (HMDA) which requires financial institutions to disclose loan-level information about mortgages: who was denied, reason for denial, etc. The disclosure of this data is intended to make home financing more transparent and facilitate the mitigation of bias relating to protected demographic variables.  Since the data is anonymized, we estimated race probabilities based on census tracts alone, i.e., we used the BISG formula but without surname probabilities.  We analyze this data across five states in the Northeast corridor: Maryland (MD), Pennsylvania (PA), Delaware (DE), New York (NY), and Virginia (VA).  We demonstrate that MICSG can provide highly accurate estimates of loan approval rates amongst four racial groups.  We benchmark our approach against the standard BISG methodology.  

\paragraph{Dataset. } We use the publicly disclosed HMDA dataset. \footnote{Data available at \color{purple}\url{https://ffiec.cfpb.gov/data-browser/}\color{black}} Across MD, PA, DE, NY, and VA, the dataset contains $n=1,459,501$ individuals.  We used a $70/30$ train-test split. In addition to the features required for BISG prediction, we use a small subset of the other features included in the dataset: loan approval (categorical), reason for denial if denied (categorical), income (continuous), and loan amount (continuous). We normalize the continuous features to have zero mean and unit variance, and one-hot encode the categorical features. We include results for the four most represented groups in the included states: Black, White, Hispanic, and Asian / Pacific Islander (API).

\paragraph{Methods.} Since the dataset is anonymized, we only use geographic information to produce BISG probabilities. To address the lack of name information, however, we compute BISG probabilities at the census tract level, which is known to be the most accurate unit of geographic division that can be used with the BISG methodology. To train MICSG we combine the BISG probabilities, with the other features we've mentioned, to produce a feature vector that contains 26 features in total.  We then use these features to estimate the likelihood that an individual belongs to each racial group using the ``new'' feature vector as input, and race as a multiclass target variable.  We use Multivariate Logistic Regression (\textbf{MR}) and Gradient Boosting (\textbf{GB}).  Multivariate Logistic Regression estimates the race distribution using a softmax function whereas Gradient Boosting estimates the race distribution using ensembles.  We compute disparity using the Bayes estimator for the MICSG method and use the weighted estimator for BISG predictions.  The true loan approval rates per group are estimated empirically from the data and are denoted under the legend \textbf{HMDA}. 

\paragraph{Results. }  We provide two sets of results.  Our first set of results, displayed in Figure \ref{fig:hmda_MICSG_vs_bisg}, shows that MICSG variants outperform BISG at estimating loan approval rates per racial group.  In this figure, dots closer to the HMDA dot, indicate accurate estimation.  We see that MICSG~(\textbf{MR}) and MICSG~(\textbf{GB}) produce the best estimates.  Additionally, BISG performs worse for racial minorities, yielding especially poor estimates for Black and Hispanic individuals.  MICSG does not appear to suffer from this limitation. 

Our second set of results visualizes the mean consistency violations of MICSG~(\textbf{MR}) compared to BISG. This plot can be interpreted as the residuals of a calibration plot over 8 bins, where the size of each dot reflects the bin size.  Our results show that MICSG performs as well as BISG for the White and Black groups, but admits far smaller mean consistency violations for the Hispanic and API groups.  We also remark that MICSG performs especially well for large bins and experiences a much smaller degradation in performance in smaller bins compared to BISG.  This effect is most noticeable in minority groups. 

\subsection*{Case Study: The Racial Composition of Political Parties in North Carolina}

To evaluate the performance of cBISG, we explore the demographics of registered voters in North Carolina.  The North Carolina voter registration file is public, making it well-suited for proxy model validation studies. 

\paragraph{Dataset.}  In our experiments, we use a randomly sampled subset of roughly ~200k voters from the North Carolina voter registration dataset. \footnote{Data available at \color{purple}\url{https://www.ncsbe.gov/results-data/voter-registration-data}\color{black}}  The dataset in its raw form contains addresses, surnames, party affiliation, and whether or not the individual voted in several past elections.  We geocoded these addresses into census tracts, dropping rows (before taking our subset) that could not be successfully geocoded.  We employed a $50/50$ train-test split. 

\paragraph{Methods. }
We compare \textbf{cBISG} to several existing BISG methods, namely, the standard \textbf{BISG} methodology, the full Bayesian Improved Surname Geocoding methodology \textbf{fBISG} from \citet{imai_addressing_2022}, and also Bayesian Improved First Name Surname Geocoding \textbf{BIFSG} \citep{voicu_using_2018}.  To compute cBISG probabilities, we used political party membership as a context variable and trained a different cBISG for each political party.  If an individual's surname was not included in the census data, we would only use race probabilities conditioned on geography, implying a uniform distribution over surname probabilities given race for that individual. Further we perform hyperparameter optimization on $\eta$ for all geographies for $\eta \in \{0, 0.1, \dots ,1\}$.   To do this optimization, we measure the estimation error for the given $\eta$ within a given geography, on the training data.  We select the hyperparameter which attained the smallest value.  We use the Bayes estimator with cBISG and the weighted estimator for the other BISG methods.  

\paragraph{Results. } Our results show that cBISG outperforms the BISG methods.  Indeed, the other approaches consistently under/overestimate the fraction of individuals in each race belonging to each political party.  Further, the direction of the estimation bias is consistent amongst non-contextual BISG methods.  We contrast this with cBISG which achieves near-perfect estimation across all races and parties, with slight estimation bias in either direction. Lastly, we remark that the estimation bias of the non-contextual proxy methods is highest for minorities. cBISG does not produce this effect.

\section*{Discussion}
In the current landscape, proxy imputation of race is a necessary tool for performing disparity estimation when race is unobserved. This is in part because of the explicit limits on collecting such data in laws like the Equal Credit Opportunity Act (ECOA) and the associated Regulation B \citep{kumar_equalizing_2022}, and in part because of laws around data protection, that have introduced difficult choices surrounding the collection and handling of sensitive data \citep{ashurst_fairness_2023}.

Our work suggests that a small amount of demographic data obtained in context can help build better proxy estimates. One approach to achieving this is to mandate purpose-limited demographic data collection. Section 4302 of the Affordable Care Act (2010) already makes headway towards this objective by requiring federal data collection to include race, ethnicity, sex, primary language, and disability status to improve the assessment of healthcare disparities. Similarly, in 2017, the Consumer Financial Protection Bureau permitted applicants to self-identify their race and ethnicity using disaggregated categories in an amendment to Regulation B. Another self-initiated approach could be for the decision-making entity to collect demographic data from customers \emph{after} decisions have been made.  This allows for such an entity to avoid the liability associated with collecting protected attributes prior to decision-making. 

\section*{Conclusion}
In this work, we study the problem of identifying racial disparities with proxy methods.  We provide two new contextual proxy algorithms, cBISG and MICSG, which leverage context to produce improved race predictions. Using these contextual proxies, we present a new disparity estimator called the Bayes estimator.  We show that if a contextual proxy satisfies mean consistency, then our Bayes estimator can be used to produce unbiased disparity estimates.  
Additionally, we present theoretical work that demonstrates what happens when a proxy is \textit{not} mean-consistent, and how mean-consistency violations affect disparity estimation.  Finally, we perform two large-scale experiments using real-world data to demonstrate the success of our proposed approaches and our proposed estimator.  

\section*{Ethical Statement}
We find it important to note that the racial categorization used by the census, and subsequently by BISG, is imperfect. This is most clearly seen in the fact that racial categories the census uses have changed many times over the years. \footnote{
\citet[Chapter 1]{ngai_impossible_2014} discusses how the census's racial categories can be fraught with politics, focusing on the infamous Johnson-Reed Act of 1924.}  Further, we acknowledge these racial categories sometimes conflate race and ethnicity and require the collapsing of distinct racial/ethnic identities into a single group.  For the purposes of identifying and mitigating racial disparities, we believe individuals should be able to self-identify their race and be ensured that this data is safely collected and maintained.

\bibliography{references}
\appendix
\onecolumn
\section{Proofs}

\paragraph{Proof of Theorem \ref{thm: unbiased}.}
\label{thm:mean-consistent-proof}
If $\omega^f_r(x,y)$ is mean consistent for all $y \in \mc Y$, then for all $r \in \mc R$ the Bayes estimator is an unbiased estimator for $\mu_f({r})$, i.e., ${\mu}_f^B(r) \rightarrow \mu_f({r})$ as $n\rightarrow \infty$.
\begin{proof}
For brevity, let $\check{f_y} = \sum_{i=1}^n \indicator[f_i = y]$. The Bayes estimator is given by 
\begin{align*}
    &\frac{\check{f}_1 \cdot \sum_{i=1}^n \omega^f_r(x_i, 1)}{\sum_{y \in \mc Y}\parens{\check{f}_y \cdot \sum_{i=1}^n \omega^f_r(x_i, y)}} \\ 
    &= \frac{\frac{1}{n}\check{f}_1 \cdot \sum_{i=1}^n \omega^f_r(x_i, 1)}{\frac{1}{n}\sum_{y \in \mc Y}\parens{\check{f}_y \cdot \sum_{i=1}^n \omega^f_r(x_i, y)}}
\end{align*}
As $n \rightarrow \infty$, this quantity converges to 
\begin{align*}
    &=\frac{\Bar{\omega}^f_r \cdot \Pr[f(X)=1]}{\sum_{y \in \mc Y}\expect\bra{\omega^f_r(x_i, y)}\Pr[f(X)=y]} 
\end{align*}
which if $\omega_r$ is mean consistent the above equals
\begin{align*}
\frac{\Pr[R =r | f(X)=1]\Pr[f(X)=1]}{\sum_{y \in \mc Y}\Pr[R =r | f(X)=y]\Pr[f(X)=y]} = \mu_f(r) 
\end{align*}
by Bayes rule. 
\end{proof}

\paragraph{Proof of Theorem \ref{thm:mean consistent_to_bias}.}
Fix $\epsilon \in \binaryrange$ and let $f \in \mc F$ be any decision function. If $\rho_r$ is a $\gamma$-biased estimate for $\theta_r$ and $\omega^f_r$ is $\paa{\frac{\epsilon\theta_r}{\nu_f} - \frac{\bar{\omega}^f_r\gamma}{\theta_r - \gamma}}$-mean consistent
then, $\mu^{B}_f(r)$ is an $\epsilon$-biased estimate for $\mu_f(r)$. 

\begin{proof}
For any $f \in \mc F$ 
\begin{align*}
    \abs{\mu_f^B(r) - \mu_f(r)} = \abs{\Pr\bra{f(X) = 1 \mid h(X) = r} - \Pr\bra{f(X) = 1 \mid R = r}}\\
\end{align*}
By Bayes rule this equals 
\begin{align*}
     \abs{\frac{\nu_f}{\rho_r} \Pr\bra{h(X) = r \mid f(X) = 1} - \frac{\nu_f}{\theta_r} \Pr\bra{R = r \mid f(X) = 1}}. 
\end{align*}
Which can be upper-bounded 
\begin{align*}
    &\le \frac{\nu_f}{\theta_r}\paa{\abs{\Pr\bra{h(X) = r \mid f(X) = 1} - \Pr \bra{R = r \mid f(X) = 1}} + \abs{\frac{\theta_r}{\rho_r} - 1}\Pr\bra{h(X) = r \mid f(X) = 1}}.
\end{align*}
Using the assumption that $\rho_r = \theta_r \pm \gamma$ we get 
\begin{align*}
    &\le \frac{\eta_f}{\theta_r}\bra{\abs{\Pr\bra{h(X) = r \mid f(X) = 1} - \Pr \bra{R = r \mid f(X) = 1}} + \paa{\frac{\theta_r}{\theta_r - \gamma} - 1}\Pr\bra{h(X) = r \mid f(X) = 1}}\\
    &\le \frac{\eta_f}{\theta_r}\bra{ \paa{\frac{\epsilon\theta_r}{\eta_f} - \frac{\gamma}{\theta_r - \gamma}\bar{\omega}^f_r} + \frac{\gamma}{\theta_r - \gamma}\bar{\omega}_r^f} = \epsilon 
\end{align*}
thus $\mu^B_f(r)$ is an $\epsilon$-biased estimate for $\mu_f(r)$ as desired. 
\end{proof}

\paragraph{Proof of Theorem \ref{thm:bias_to_consistent}. }
If for any $f \in \mc F$ it holds that $\mu_f^B(r)$ is an $\epsilon$-biased estimate for $\mu_f(r)$ then $\omega^f_r$ is $\paa{ \frac{\epsilon \theta_r}{\eta_f} + \gamma \frac{ \mu^B_f(r)}{\eta_f}}$ mean consistent.
\begin{proof}
Pick $f \in \mc F$ and let $\Pr[f(X) = 1 \mid h(X) = r]$ be the $\epsilon$-biased estimate.  We apply Bayes rule to obtain
\begin{align*}
      \abs{\bar{\omega}^f_r - \phi_r} & = \frac{\theta_r}{\nu_f}\abs{\frac{\rho_r}{\theta_r}\Pr[f(X) = 1 \mid h(X) = r] - \mu_f(r)}. \\ 
\end{align*}
and then we upper bound the above as
\begin{align*}
    &\le \frac{\theta_r}{\nu_f}\paa{\abs{\Pr[f(X) = 1 \mid h(X) = r] - \mu_f(r)} + \Pr[f(X) = 1 \mid h(X) = r] \abs{1 - \frac{\rho_r}{\theta_r}}}\\
    &\le \frac{\theta_r}{\eta_f}\paa{\epsilon + \mu^B_f(r) \frac{\gamma}{\theta_r}} = \frac{\epsilon \theta_r}{\eta_f} + \gamma \frac{\mu^B_f(r)}{ \eta_f} \\
\end{align*}
whence we find that $\omega^f_r$ is $\paa{\frac{\epsilon \theta_r}{\eta_f} + \gamma \frac{\mu^B_f(r)}{\eta_f}}$-mean consistent as desired.
\end{proof}

\end{document}